\pdfoutput=1

\documentclass[aip,jcp,reprint]{revtex4-1} 

\usepackage{graphicx}
\usepackage{xcolor}
\usepackage{calc}

\usepackage{titlesec}
\titlespacing\section{0pt}{6pt plus 2pt minus 2pt}{6pt plus 2pt minus 2pt}
\titlespacing\subsection{0pt}{4pt plus 2pt minus 2pt}{4pt plus 2pt minus 2pt}

\usepackage{enumitem}
\newlist{enumerateinl}{enumerate*}{1}
\setlist*[enumerateinl,1]{%
  label=(\alph*),
}

\usepackage[colorlinks=true,linkcolor=blue,urlcolor=blue,citecolor=blue,bookmarks=false]{hyperref}

\newcommand{\highlight}[1]{\textbf{#1}}

\makeatletter
	\def\Dated@name{Submitted to \textit{The Journal of Chemical Physics}}
\makeatother

\begin{document}

\title{Guest Editorial: Special Topic on Software for Atomistic Machine Learning}

\author{Matthias Rupp}
\email[]{mrupp@mrupp.info}
\affiliation{\begin{minipage}{\linewidth}\small Luxembourg Institute of Science and Technology, L-4362 Esch-sur-Alzette, Luxembourg\end{minipage}}

\author{Emine K{\"u}{\c{c}}{\"u}kbenli}
\email[]{ekucukbenli@nvidia.com}
\affiliation{\begin{minipage}[t]{0.8\linewidth}\small John A. Paulson School of Engineering and Applied Sciences, Harvard University, Cambridge, Massachusetts 02138, USA, and, Nvidia Corporation, Santa Clara, CA, USA \end{minipage}}

\author{G{\'a}bor Cs{\'a}nyi}
\email[]{gc121@cam.ac.uk}
\affiliation{\begin{minipage}[t]{0.8\linewidth}\small Engineering Laboratory, University of Cambridge, Cambridge, CB2 1PZ, United Kingdom\end{minipage}}

\date{}

\maketitle

\onecolumngrid
\vspace*{-2\baselineskip}
\begin{center}
A survey of the contributions to the Special Topic on Software for Atomistic Machine Learning.
\end{center}
\bigskip
\twocolumngrid


\section{Introduction}

\noindent
Welcome to the Journal of Chemical Physics' Special Topic on Software for Atomistic Machine Learning.
For some years now, search engines have been dominating our online experience and have essentially overtaken libraries, whether physical or digital, as the means to find information we are looking for.
Most readers of an original research article find it by citation or direct search, and not by browsing journal volumes. 
Given this, one might wonder what the utility of a Special Topic issue of a scientific journal might be. 

However, publishing papers on scientific software has traditionally been somewhat neglected, with few go-to journals for publishing, such as the Journal of Open Source Software or Computer Physics Communications. Typical published software papers tend to discuss relatively mature software packages. 
In this context, the Journal of Chemical Physics' initiative~\cite{smcm2021} to support software publications is especially welcome.
Given the huge activity currently taking place across many sub-fields and communities in new software development for atomistic machine learning (ML), this landscape is changing fast. 
Arguably, many papers in this Special Topic issue might not have been written if it were not for the impetus provided by this Special Topic issue. 

Beyond their individual value regarding specific software packages, these papers as a collection provide a snapshot at this moment in time of the kinds of tools that people use and the goals they set themselves and achieve for the software implementations of their methods.
Table~\ref{tab:overview} presents an overview of the 28 invited and contributed articles.~\cite{crh2022q,fwsa2022q,bfl2022q,hqlb2023q,kpjc2023q,shlg2023q,lama2023q,lhtr2023q,btrp2023q,sch2023q,skvm2023q,gwjs2023q,td2023q,gts2023q,secd2023q,vrm2023q,zzzw2023q,bfbc2023q,pgsn2023q,plsk2023q,sshn2023q,psg2023q,gwcb2023q,sjwp2023q,vfmt2023q,woco2023q,lfk2023q,kdcb2023q}
Of these, 18 (64\,\%) deal directly with machine-learning interatomic potentials (MLIPs).
The other 10 articles cover a broad range of subjects, ranging from sampling to data set repositories and workflows. 

In the following, we give an overview of these contributions.

\vfill\eject


\section{Contributions}

\noindent%
Since their beginnings in the 1980s and 1990s, MLIPs have undergone tremendous development and now constitute a highly active field of research.
Some modern MLIPs can predict forces with an accuracy close to the underlying ab-initio reference method for atomistic systems with many chemical elements and millions of atoms while still providing orders of magnitude of acceleration.
These capabilities have increasingly enabled scientific applications using  MLIPs that would not otherwise have been possible.

Consequently, there is an trend to  \highlight{directly integrate MLIPs into molecular dynamics codes}.
In this Special Topic, four contributions describe the integration of
\begin{enumerateinl}
    \item neuro-evolution potentials into \texttt{GPUMD} (Graphics Processing Units Molecular Dynamics), including improved featurization, GPU code, active learning, and supporting Python packages \texttt{gpyumd}, \texttt{calorine}, and \texttt{pynep};~\cite{fwsa2022q}
    \item PhysNet into \texttt{CHARMM} (Chemistry at HARvard Macromolecular Mechanics) via a new \texttt{MLpot} extension of the \texttt{pyCHARMM} interface, with para-chloro-phenol as an example;~\cite{skvm2023q}
    \item general MLIPs into \texttt{CASTEP} (CAmbridge Serial Total Energy Package), including active learning, using the example of a GAP/SOAP model;~\cite{secd2023q}
    \item ephemeral data-derived potentials with \texttt{AIRSS} (Ab Initio Random Structure Search).~\cite{sjwp2023q}
\end{enumerateinl}

An important milestone in MLIP development was the introduction of \highlight{artificial neural network MLIPs} that were able to efficiently handle large (``high-dimensional'') atomistic systems by Behler and Parrinello~\cite{bp2007q}.
Six contributions in this Special Topic present MLIPs related to Behler-Parrinello networks:
\begin{enumerateinl}
    \item neuro-evolution potentials;~\cite{fwsa2022q}
    \item the atomic energy network ({\ae}net) in a \texttt{PyTorch} implementation for GPU support;~\cite{lama2023q}
    \item deep potentials via \texttt{DeePMD-kit}, recent improvements including attention-based features, learning dipoles and polarizabilities, long-range interactions, model compression, and GPU acceleration;~\cite{zzzw2023q}
    \item multi-layer-perceptron-based MLIPs via \texttt{PANNA} (Properties from Artificial Neural Network Architectures), with improved GPU support based on TensorFlow and long-range electrostatic interactions through a variational charge equilibration scheme; \cite{plsk2023q}
    \item ephemeral data-derived potentials (EDDPs) for atomistic structure prediction, including ensemble-based uncertainties.~\cite{sjwp2023q}
\end{enumerateinl}


\begin{table*}[htbp]
	\caption{%
            \emph{Overview of contributions} to the Special Topic.
            See Table~\ref{tab:glossary} for acronyms.
            \label{tab:overview}
        }

	\smallskip\small

	\begin{ruledtabular}
        \begin{tabular}{%
            @{}%
            p{6mm}%
            p{22mm}%
            p{8mm}%
            p{21mm}%
            p{12mm}
            p{42mm}%
            p{\linewidth-6mm-22mm-8mm-21mm-12mm-42mm-12\tabcolsep}%
            @{}}
            
            Ref. & Package & ML & Language & License & Data & Comments \\ \hline

            \citenum{crh2022q} & \raggedright\texttt{AGOX}
                & ---  & Python   & GPL3     & Pt$_{14}$/Au(100) 
                & ASE; structure search \tabularnewline

            \citenum{fwsa2022q} & \raggedright\texttt{GPUMD}
                & ANN & C++ & GPL3     & Si, C, MD17 
                & CUDA; neuro-evolution potentials \tabularnewline

            \citenum{bfl2022q} & \raggedright\texttt{QML-lightning}
                & GPR & Python  & MIT & QM9, MD17, 3PBA 
                & PyTorch; FCHL19, GPR, RFF, SORF \tabularnewline

            \citenum{hqlb2023q} & \raggedright\texttt{PESPIP}
                & LR  & \raggedright Mathematica C++, Fortran Perl, Python & --- & H$_2$O, hydrocarbons 
                & permutationally invariant polynomials \tabularnewline

            \citenum{kpjc2023q} & \raggedright---
                & GPR & ---       & ---        & Pt nanoparticles 
                & GAP/SOAP application \tabularnewline

            \citenum{shlg2023q} & \raggedright\texttt{SchNetPack2}
                & ANN  & Python & MIT & QM9, MD17, C$_2$H$_6$O 
                & PyTorch; MD, FieldSchnet, PaiNN \tabularnewline

            \citenum{lama2023q} & \raggedright\texttt{{\ae}net-\\PyTorch}
                & ANN & Python & MIT & \raggedright TiO$_2$, LiMoNiTiO amorphous Li$_x$Si 
                & PyTorch; atomic energy network \tabularnewline

            \citenum{lhtr2023q} & \raggedright\texttt{DScribe}
                & ---  & Python, C++  & Apache2 & \raggedright CsPb(Cl/Br)$_3$, Cu clusters
                & atomistic featurization \tabularnewline 

            \citenum{btrp2023q} & \raggedright\texttt{mlcolvar}
                & DR & Python & MIT & \raggedright alanine dipeptide, aldol reaction, chignolin 
                & \raggedright dimensionality reduction, collective variables, enhanced sampling \tabularnewline 

            \citenum{sch2023q}  & \raggedright\texttt{AGOX}
                & --- & Python & GPL3 & \raggedright rutile SnO$_2$(110)-(4$\times$1), olivine (Mg$_2$SiO$_4$)$_4$ 
                & ASE; structure generation \tabularnewline

            \citenum{skvm2023q} & \raggedright\texttt{CHARMM}
                & ANN  & Python & --- & para-chloro-phenol 
                & PhysNet integration \tabularnewline

            \citenum{gwjs2023q} & \raggedright\texttt{AL4GAP}
                & GPR    & Python & MIT & molten salts
                & ensemble active learning, GAP \tabularnewline 

            \citenum{td2023q} & \raggedright\texttt{XPOT}
                & BO  & Python & GPL2 & Si, Cu, Si, a~C 
                & hyperparameter optimization \tabularnewline

            \citenum{gts2023q} & \raggedright\texttt{PES-Learn}
                & ANN & Python & BSD3 & \raggedright methanol, HCOOH 
                & benchmark multifidelity approaches \tabularnewline

            \citenum{secd2023q} & \raggedright\texttt{CASTEP}
                & GPR  & Python & CAL & Al$_2$O$_3$, Si, a-C 
                & active-learning MLIP for CASTEP \tabularnewline

            \citenum{vrm2023q}  & \raggedright\texttt{q-pac}
                & GPR  & Python & MIT & QM9, ZnO, ZnO$_2$ 
                & kernel charge equilibration \tabularnewline 

            \citenum{zzzw2023q} & \raggedright\texttt{DeePMD-kit}
                & ANN  & \raggedright Python, C/C++ & LGPL3 & \raggedright H$_2$O, Cu, HEA, OC2M, SPICE 
                & TensorFlow; deep potentials \tabularnewline

            \citenum{bfbc2023q} & \raggedright\texttt{sphericart}
                & ---    & C++, Python & Apache2 & --- 
                & PyTorch; fast spherical harmonics \tabularnewline 

            \citenum{pgsn2023q} & \raggedright\texttt{MLIP-3}
                & NLR  & C++ & BSD & Cu(111) 
                & moment tensor potentials \tabularnewline 

            \citenum{plsk2023q} & \raggedright\texttt{PANNA2}
                & ANN  & \raggedright Python & MIT & \raggedright rMD17, C, NaCl clusters 
                & TensorFlow; ANN MLIP training \tabularnewline

            \citenum{sshn2023q} & \raggedright\texttt{DeepQMC}
                & ANN & Python & MIT & \raggedright NH$_3$, CO, N$_2$, cyclobutadiene, reactions; ScO, TiO, VO, CrO 
                & \raggedright JAX; variational quantum Monte Carlo \tabularnewline 

            \citenum{psg2023q} & \raggedright\texttt{SISSO++}
                & SR   & C++, Python & Apache2 & --- 
                & symbolic regression \tabularnewline

            \citenum{gwcb2023q} & \raggedright\texttt{wfl}, \texttt{ExPyRe}
                & --- & Python & \raggedright GPL2 & --- 
                & ASE-based workflows \tabularnewline 

            \citenum{sjwp2023q} & \raggedright\texttt{EDDP}
                & ANN & Fortran, Julia & \raggedright GPL2, MIT & \raggedright C, Pb, ScH$_{12}$, Zn(CN)$_2$ 
                & ephemeral data-derived potentials \tabularnewline

            \citenum{vfmt2023q} & \raggedright\texttt{ColabFit Exchange}
                & ---   & --- & --- & many datasets 
                & dataset repository \tabularnewline

            \citenum{woco2023q} & \raggedright\texttt{ACEpoten-\\tials.jl}
                & LR  & Julia & MIT & \raggedright 6 elements, H$_2$O, AlSi$_{10}$, polyethylene glycol, CsPbBr$_3$ 
                & linear GPR/ACE MLIPs \tabularnewline 

            \citenum{lfk2023q}  & \raggedright\texttt{glp}
                & AD  & Python & MIT & SnSe 
                & JAX; auto-differentiation, heat flux \tabularnewline

            \citenum{kdcb2023q} & \raggedright\texttt{QUIP}
                & GPR & \raggedright Fortran, Python, C & \raggedright GPL2, ASL & \raggedright Si, core e binding energies, MoNbTaVW 
                & GAP MLIPs, MPI parallelization \tabularnewline 

        \end{tabular}
    \end{ruledtabular}
\end{table*}


\begin{table}[htbp]

    \vspace*{-8pt}
    \caption{%
        \emph{Glossary.}
        \label{tab:glossary}
    }

    \smallskip\small
    
    \begin{ruledtabular}
    \begin{tabular}{@{}p{5em}p{\linewidth-2\tabcolsep-5em}@{}}

        Acronym & Meaning \\
        \hline

    a-C    & Amorphous Carbon \tabularnewline
    AD     & Automatic Differentiation \tabularnewline
    ACE    & Atomic Cluster Expansion \tabularnewline
    AGOX   & Atomistic Global Optimization~X \tabularnewline
    AIRSS  & Ab Initio Random Structure Search \tabularnewline
    AL     & Active Learning \tabularnewline
    ANN    & Artificial Neural Network \tabularnewline
    ASE    & Atomic Simulation Environment \tabularnewline
    ASL    & Academic Software License \tabularnewline
    BO     & Bayesian Optimization \tabularnewline
    BSD    & Berkeley Software Distribution \tabularnewline
    CAL    & CASTEP Academic License \tabularnewline
    CASTEP & CAmbridge Serial Total Energy Package \tabularnewline
    CHARMM & \raggedright Chemistry at HARvard Macromolecular Mechanics \tabularnewline
    CUDA   & Compute Unified Device Architecture \\
    DR     & Dimensionality Reduction \tabularnewline
    EDDP   & Ephemeral Data-Derived Potential \tabularnewline
    FCHL19 & Faber, Christensen, Huang, Lilienfeld 2019 \tabularnewline
    GAP    & Gaussian Approximation Potential \tabularnewline
    GOFEE  & \raggedright Global Optimization with First-principles Energy Expressions \tabularnewline
    GPL    & General Public License \tabularnewline
    GPR    & Gaussian Process Regression \tabularnewline
    GPU    & Graphics Processing Unit \tabularnewline
    GPUMD  & Graphics Processing Units Molecular Dynamics \tabularnewline
    JAX    & Just After eXecution \tabularnewline
    LGPL   & Lesser General Public License \tabularnewline
    LR     & Linear Regression \tabularnewline
    MIT    & Massachusetts Institute of Technology \tabularnewline
    MD     & Molecular Dynamics \tabularnewline
    ML     & Machine Learning \tabularnewline
    MLIP   & Machine-Learning Interatomic Potential \tabularnewline
    MIP    & Message-Passing Interface \tabularnewline
    NLR    & Non-Linear Regression \tabularnewline
    PaiNN  & Polarizable atom interaction Neural Network \tabularnewline
    PANNA  & \raggedright Properties from Artificial Neural Network Architectures \tabularnewline
    PES    & Potential Energy Surface \tabularnewline
    PIP    & Permutationally Invariant Polynomial \tabularnewline
    PLUMED & PLUgin for MolEcular Dynamics \tabularnewline
    QML    & Quantum Machine Learning \tabularnewline
    QUIP   & QUantum mechanics and Interatomic Potentials \tabularnewline
    RFF    & Random Fourier Features \tabularnewline
    SISSO  & \raggedright Sure Independence Screening and Sparsifying Operator \tabularnewline
    SOAP   & Smooth Overlap of Atomic Positions \tabularnewline
    SORF   & Structured Orthogonal Randomized Features, \tabularnewline
    SR     & Symbolic Regression \tabularnewline
    XPOT   & Cross-Platform Optimizer for Potentials \tabularnewline
    \end{tabular}
    \end{ruledtabular}
\end{table}

\highlight{Message-passing neural networks} allow the exchange of information between atoms beyond their local environments by repeatedly passing messages between them.
Two contributions provide such MLIPs:
\begin{enumerateinl}
    \item the \texttt{SchNetPack2} library provides improved support functionality, including data sparsity, equivariance, and PyTorch-based MD, and provides four MLIPs: SchNet, FieldSchnet (external fields), as well as PaiNN and SO3net (equivariance);~\cite{shlg2023q}
    \item the existing PhysNet MLIP is integrated into \texttt{CHARMM}.~\cite{skvm2023q}
\end{enumerateinl}

\highlight{Kernel-based learning} is another ML approach that many MLIPs employ.
Gaussian process regression (GPR), in particular, has been frequently used since the introduction of Gaussian Approximation Potentials (GAP)~\cite{bpkc2010q}.
Five contributions in this Special Topic deal with kernel-based MLIPs:
\begin{enumerateinl}
    \item the \texttt{QML-lightning} package provides GPU-accelerated sparse approximate GPR and representations (random features, FCHL19);~\cite{bfl2022q}
    \item \texttt{AL4GAP} provides ensemble-based active learning for GAP MLIPs to study charge-neutral molten-salt mixtures.~\cite{gwjs2023q} 
    \item the \texttt{q-pac} package implements kernel charge equilibration based on sparse GPR for long-range electrostatic interactions, non-local charge transfer, and energetic response to external fields;~\cite{vrm2023q}
    \item the \texttt{QUIP} package allows training and deployment of GAP models, recent additions including distributed training via the Message Passing Interface (MPI) and compressed features;~\cite{kdcb2023q}
    \item an application study develops a GAP/SOAP MLIP to obtain the pressure-temperature phase diagram of Pt and to simulate the spontaneous crystallization of a large Pt nanoparticle.~\cite{kpjc2023q}
\end{enumerateinl}

\highlight{Other} ML approaches can be used to develop MLIPs, notably linear regression.
Three contributions describe such MLIPs:
\begin{enumerateinl}
    \item the \texttt{ACEpotentials.jl} Julia package provides MLIPs based on linear GPR and the atomic cluster expansion (ACE) representation, including uncertainties and active learning;~\cite{woco2023q}
    \item the \texttt{PESPIP} package provides MLIPs based on permutationally invariant polynomials in Morse-transformed interatomic distances, including their optimization;~\cite{hqlb2023q}
    \item the \texttt{MLIP-3} package provides moment tensor potentials, including fragment-based active learning for large simulation cells.~\cite{pgsn2023q} 
\end{enumerateinl}

Besides the MLIPs themselves, seven contributions provide \highlight{auxiliary tooling and analysis} that focus on specific aspects of MLIPs but are not specific to one MLIP:
\begin{enumerateinl}
    \item the \texttt{DScribe} library provides many atomistic representations, and has been extended to include Valle-Oganov materials fingerprints and derivatives for all representations;~\cite{lhtr2023q}
    \item the \texttt{sphericart} package implements efficient real-valued spherical harmonics, a key ingredient of many representations for MLIPs, including stable Cartesian derivatives;~\cite{bfbc2023q}
    \item the \texttt{XPOT} (Cross-Platform Optimizer for Potentials) package provides hyperparameter optimization for MLIPs;~\cite{td2023q}
    \item \texttt{PES-Learn} benchmarks four approaches to train neural-network MLIPs on data with different levels of fidelity (e.g., low and high accuracy);~\cite{gts2023q}
    \item the \texttt{wfl} (Workflow) and \texttt{ExPyRe} (Execute Python Remotely) packages provide workflow management routines tailored for atomistic simulations and MLIP development;~\cite{gwcb2023q}
    \item the ColabFit Exchange repository hosts hundreds of diverse datasets of atomistic systems in extended XYZ format for MLIP benchmarking and development.~\cite{vfmt2023q} 
    \item the \texttt{glp} package demonstrates how to use automatic differentiation to efficiently obtain forces, stress, and heat flux for message-passing MLIPs;~\cite{lfk2023q}
\end{enumerateinl}

The remaining five contributions span a wide range of \highlight{other areas}:
\begin{enumerateinl}
    \item the \texttt{mlcolvar} library implements multiple dimensionality reduction methods to identify collective variables for analysis and enhanced sampling in MD simulations, including an interface to the PLUMED (PLUgin for MolEcular Dynamics) software;~\cite{btrp2023q} 
    \item the \texttt{AGOX} (Atomistic Global Optimization~X) package enables developing global optimization algorithms for atomistic structure search, including random search, basin hopping, evolutionary algorithms, and global optimization with first-principles energy expressions (GOFEE);~\cite{crh2022q}
    \item the same \texttt{AGOX} package also includes structure generation based on local optimization in ``complementary energy'' landscapes (oversmoothed MLIPs), favoring structures with fewer distinct local motifs;~\cite{sch2023q}
    \item the \texttt{DeepQMC} package provides a framework for neural network-based variational quantum Monte Carlo methods, including PauliNet, FermiNet, and DeepErwin.~\cite{sshn2023q}
    \item the \texttt{SISSO++} (Sure Independence Screening and Sparsifying Operator) software offers symbolic regression, including recent improvements in expression representation, support for units, nonlinear parametrization, and the solver algorithm.~\cite{psg2023q}
\end{enumerateinl}
 

\section{Summary}

This Special Topic on Software for Atomistic Machine Learning contains 28 invited and contributed articles.
They range from MLIPs based on neural networks, kernel models, and linear regression, as well as their integration into MD codes, to auxiliary tooling, structure search, dimensionality reduction, quantum Monte Carlo methods, and symbolic regression. We hope you enjoy reading this community effort at capturing the state of the field at this moment. 

The Journal of Chemical Physics encourages and welcomes submissions of original articles describing software implementations relevant to the broad remit of the journal. 

\bigskip


\end{document}